# [1]INTERNATIONALIZATION OF MALAYSIAN MATHEMATICAL AND COMPUTER SCIENCE JOURNALS

**A.N. Zainab**
Library & Information Science Unit,
Faculty of Computer Science & Information Technology,
University of Malaya, Kuala Lumpur, Malaysia
e-mail: zainab@um.edu.my

**ABSTRACT**

The internationalization characteristics of two Malaysian journals, *Bulletin of the Malaysian Mathematical Sciences Society* (indexed by ISI) and the *Malaysian Journal of Computer Science* (indexed by Inspec and Scopus) is observed. All issues for the years 2000 to 2007 were looked at to obtain the following information, (i) total articles published between 2000 and 2007; (ii) the distribution of foreign and Malaysian authors publishing in the journals; (iii) the distribution of articles by country and (iv) the geographical distribution of authors citing articles published in the journals. Citation to articles is derived from information given by Google scholar. The results indicate that both journals exhibit average internationalization characteristics as they are current in their publications but with between 19% -30% international composition of reviewers or editorials, publish between 36%-79% of foreign articles and receive between 60%-70% of citations from foreign authors.

## INTRODUCTION

Kieling and Goncalves (2007) highlighted the problems affecting scientific journals from developing countries such as their limited accessibility and visibility. In their context, it was journals in Psychiatry from Brazil in particular and journals which carry Portuguese and Spanish articles in general. They then go on to describe how the Brazilian journal "*Revista Brasileira de Psiquitria*" (RBP) struggled, to get itself indexed by Medline in 2003 and the ISI in 2005. The authors also noted that the inclusion of RBP in the databases has resulted in an increase in the number of submissions, and a gentle rise of the journal's impact factor value. To further increase RBP's visibility, the journal has also opted for an online manuscript

---

[1] Improved version of the paper presented at the Workshop on Managing Scholarly Journals, 13-16 January 2008, Seaview Hotel, Langkawi, Malaysia

submission system to speed-up the peer review process and increase accuracy. Another problem faced by authors in developing countries is getting their works accepted in leading journals in the various fields. Patel and Sumathipala (2001) revealed authors from developing countries contributed only 6% of articles published in six major European and American psychiatric journals between 1996 and 1998. This poor performance of authors in developing countries was also noted by Catapano and Castle (2003). The reason for this low acceptance may be due to the poor representation in editorial and advisory boards of leading journals and the lack of interest of research reported from developing countries.(Saxena et al. 2003).

There has been a growth of studies in recent years emanating from either non-English speaking countries in the various disciplines or editorial members of journals in a specific discipline in an attempt to identify and understand the internationalization characteristics. Gutierrez and Lopez-Nieva (2001) from Universidad Complutense de Madrid in Spain when studying to which extent international journals in human geography are really international observed that the discipline itself is more national or regional in nature. Yamazaki and Zhang (1997) from Japan analysed the characteristics of the internationalization of four English language journals published in Japan in life sciences and revealed that three of the journals were not international in terms of their impact factors and the geographic distribution of authors. Calafat, Becona and Bobes (2003) lamented on the difficulties faced by the 15-year old Spanish language journal, *Addicciones* in getting itself indexed into Medline and in enticing active authors to contribute. Wang, Wang and Weldon (2007) analysed ten China's English language scientific journals and found that even though the journals may perform reasonably well in terms of impact factor, they do not exhibit sufficient internationalization characteristics as they publish less than 20% of foreign articles. In other words the Chinese journals are publishing more articles by Chinese authors as well as citing Chinese articles and this makes them more national rather than international. In these cases, impact factor counts may be pushed up by the sheer number of citations from Chinese nationals.

In this study, internationalization focuses on the composition of authors and citers. Internationalization in this paper is defined as integrating an international, global dimension into the delivery of scholarly journals in Malaysia. The product delivered would be international in characteristics and consumed by an internationally-based market.  This working definition is taken from Knight (2003) who defines and use the term in an educational context. Internationalization connotes a changing state and not something static. The concept refers to the delivery of a product which reflects an international dimension. This international dimension remains central, not marginal, and ensures the journals sustainability. The product delivered would reflect an entity comprising internationally contributed articles, which in turn are internationally consumed or cited. Internationalization is a part of globalization. Internationalization should change the environment of journal publishing and globalization is changing the world of internationalization. In fact, Malaysian journals should try to internationalize themselves first and globalization will then takes its own course.

**METHODOLOGY**

This paper adopted two of the factors used by Wormell (1998) who studied the internationalization characteristics of 7 journals in the field of Library and Information Science (LIS). The aim is to answer the following questions: (a) Where does most of the intellectual input come from?, and (b) In which regions are the users of the journals concentrated? The first involved analyzing the geographical distribution pattern of authors publishing in the selected journals. The second involved looking at who are citing articles published in these journals. These two variables were also proposed by Hjortgaard Christensen and Ingwersen (1996) as indicators for the internationality of core journals in library and information science.

This study will compare two Malaysian journals, one is indexed by the ISI database and the other is indexed by Scopus. Since both journals chosen are covered by the databases, it is assumed that they should exhibit some of the characteristics of internationality. The journals are *Bulletin of the Malaysian Mathematical Sciences Society* (BMMSS) and the *Malaysian Journal of Computer Science* (MJCS). The journals are selected for convenience as both journals are indexed by *MyAIS (Malaysian Abstracting and Indexing System*) (http://myais.fsktm.um.edu.my) and accessible over the web.  A longitudinal approach is taken as the information about articles published in both journals can be obtained from MyAIS, which includes the affiliations and country of contributing authors. All issues for the years 2000 to 2007 were looked at to obtain the following information, (i) total articles published between 2000 and 2007; (ii) the distribution of foreign and Malaysian authors publishing in the journals; (iii) the distribution of articles by country; and (iv) the geographical distribution of authors citing articles published in the journals. Citation to articles is derived from information given by Google scholar. This is because citation information for BMMSS and MJCS is not available from the ISI and Scopus as both have only begun to be indexed by the ISI and Scopus respectively in 2007. Google scholar has been indexing MJCS through MyAIS since 2006 and some citation count is expected to be available.

**THE JOURNALS AND TOTAL ARTICLES PUBLISHED**

The *Bulletin of the Malaysian Mathematical Society* (BMMSS) (ISSN 0126-6705) is published by the Malaysian Mathematical Sciences Society, which is currently hosted by the School of Mathematical Sciences, Universiti Sains Malaysia (USM). This journal is published twice a year. It is not accessible online full-text but does provide access to its tables of contents and abstracts for 19 of its issues from Volume 21, no.1 1998 to Volume 30, no. 1 2007. This is provided through its website at http://math.usm.my/bulletin/html/research.htm. The website also provides instructions to authors and information about its reviewing members. The journal adopts the publishing style of the American Mathematical Society. Articles are accepted both in English and the Malay language. The journal charges for additional reprints ordered by authors. The editorial board members comprise mainly Malaysian nationals from various academic institutions. The Editor in Chief is from the School of Mathematical Sciences in USM, and the other 16 reviewing members comprise 5 from USM, 1 from Universiti Kebangsaan Malaysia, 1 from the University of Malaya, 1 from Universiti Teknologi Mara Malaysia, 1 from Universiti Teknologi Malaysia, 1 from Universiti Putra Malaysia, 2 from India (Department of Mathematics, Indian Institute of Technology Madras Chennai, India and Department of Mathematics, University of Delhi) and 1 each from the National Tsing Hua University, Hsinchu, Taiwan; Gunma University Kiryu, Japan; Institut Teknologi Bandung Indonesia and Lincoln University Canterbury New Zealand. The reviewing board composition is therefore partially international with about 37% foreign representation. This journal is being indexed in the ISI *Science Citation Index* beginning 2007.

The *Malaysian Journal of Computer Science* (MJCS) (ISSN 0127-9084) is published by the Faculty of Computer Science and Information Technology, University of Malaya. This is a typically faculty-based journal in Malaysia where the editorial members are mainly academics from the faculty that publishes it. The reviewers for journals like this would comprise a few foreign personals. MJCS is published twice a year in June and December. This journal follows the hybrid publishing model, electronically at http://ejum.fsktm.um. edu.my which is freely accessible to users and subscription is charged for those who require print copies. This journal is indexed by Scopus and Inspec. The reviewers comprise 30 academics from Malaysian Universities, 1 each from France, India and the USA and 2 each from England and Hong Kong. There is low representation of foreigners (19%) in the journal's reviewing board and therefore does not reflect high internationality.

Table 1 indicates the total number of papers published by both journals from 2000 to 2007. The table shows that both journals are current in the delivery of issues. BMMSS for instance seems to even increase its total published articles especially from 2003 onwards. This indicates a healthy pool of publishable submissions, a healthy backlog of manuscripts and an assurance for longevity of the journals. Both journals also indicated a stable editorial practice of publishing a constant number of papers per year, which may have been controlled by the cost of paper printing.

Table 1: Number of papers published in BMMSS and MJCS, 2000 to 2007

| Year | BMMSS | MJCS | Total |
|------|-------|------|-------|
| 2007 | 21 | 16 | 37 |
| 2006 | 23 | 16 | 39 |
| 2005 | 21 | 15 | 36 |
| 2004 | 24 | 16 | 40 |
| 2003 | 22 | 16 | 38 |
| 2002 | 16 | 18 | 34 |
| 2001 | 18 | 20 | 38 |
| 2000 | 19 | 18 | 37 |
| **TOTAL** | **164(54.8%)** | **135(45.1%)** | **299(100%)** |

**THE DISTRIBUTION OF FOREIGN AND MALAYSIAN AUTHORS**

The composition of contributing authors in a journal is felt to be an important characteristic in the internationalization of journals. The internationalization of journals in terms of its contributing authors is well discussed in literature. Elster and Chen (1994) studied international contributions to the *American Journal of Roentgeology* (AJR) and indicated an increment of international submissions and a decrease in the contributions from the United States and Canada. Roger (2001) and Jenkins (2001) verified this situation. Jenkin found that in 2000, AJR received more submissions from international authors than from those residing in the United States. Ozsunar et al (2001) also indicated this condition for the journal *Radiology*, which indicated 405 contributions from international authors for articles published in 1999. Chen, Jenkin and Elster (2003) revisited the *Journal of Roentgenology* and looked at articles submitted between 2000 and 2002 and found that international contributions amount to 37% (602/1624).

Table 2 indicates the distribution of international authors compared to Malaysian authors based on their affiliation status. The results indicate that BMMS publishes a higher proportion of submission from foreign authors, whereas MJCS publishes more Malaysian articles. This may infer that as a journal moves towards internationalization, its presence began to be felt by foreign authors who would be encouraged to choose the journal for their submissions. On the other hand, the ISI may be more inclined to cover journals with a higher mix of international to national articles and in so doing makes the journal a target for submission for authors whose institutions regard the ISI journals are of higher value in promotion exercises. It is not surprising that an ISI covered journals tend to receive and publish more international than national contributions.

Table 2: Foreign and Malaysian contributions in BMMSS and MJCS

| Year | BMMS | | MJCS | | Total | |
|------|------|------|------|------|------|------|
| | International | Malaysian | International | Malaysian | International | **Malaysian** |
| **2007** | 21 | 0 | 8 | 8 | 29 | **8** |
| **2006** | 18 | 5 | 6 | 10 | 24 | **15** |
| **2005** | 17 | 4 | 5 | 10 | 22 | **14** |
| **2004** | 23 | 1 | 3 | 13 | 26 | **14** |
| **2003** | 11 | 11 | 7 | 9 | 18 | **20** |
| **2002** | 13 | 3 | 10 | 8 | 23 | **11** |
| **2001** | 14 | 4 | 7 | 13 | 21 | **17** |
| **2000** | 13 | 6 | 2 | 16 | 15 | **22** |
| **Total** | **130 (79%)** | **34 (21%)** | **48 (36%)** | **87 (64%)** | **178 (60%)** | **121 (40%)** |

The total number of foreign articles (178) is then tabulated, to find out the country of contributions and the results is displayed in Figure 1. Most of the contributions come from the East Asia and Asia-Pacific region, with the majority are from India, followed by China, Turkey, Asians residing in the USA, Japan, Iran and Indonesia. There are 26 countries contributing one publication each and these contributors come from various parts of the world such as, Mexico, Finland, Serbia, South Africa and to the east, Korea. This indicates the internationalization character of the two journals especially, BMMSS, which publishes the bulk of foreign articles.

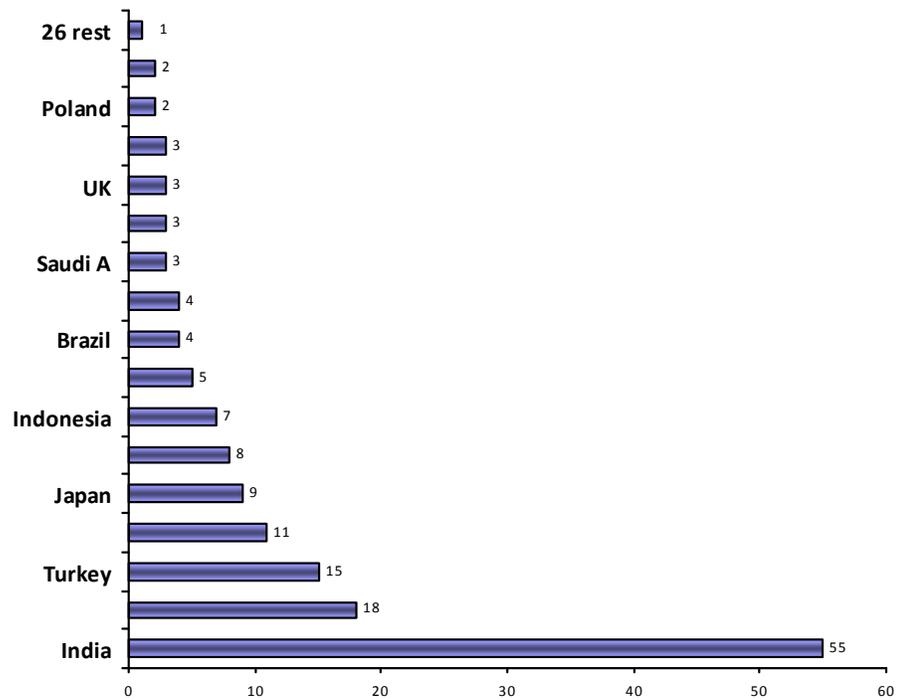

Figure 1: Country distribution of articles published in BMMSS and MJLS

The relationships between an increase in the number of international contributions to a journal when it reaches a certain standard is also indicated by Sin (2005) who analysed the geographical affiliations of authors in 20 international library and information science (LIS) journals published between 1980 and 2003. The results are obtained from a total of 12,511 research papers. The findings found an increase in the internationalization of LIS authorship over the years. Authors from the USA account for over 56% (9735) of contributions, followed by the UK (10%), Canada (4%), Netherlands, France, Japan and Germany (roughly 2% respectively). This shows that USA and UK contributed to 64% of all articles and this pattern of authorship is still unevenly skewed towards the developed and economically strong nations. This steady increase in the number of submissions as well an increase in submissions from foreign authors was also observed by Kieling and Goncalves for the Brazilian journal *Revista Brasileira de Psiquiatria*. They observed that over 30% of the authors who cited articles from RBP came from the United States, England and Canada but there was still high citations received from Brazilian authors. They also noticed that most of the articles cited by non-Brazilians were written in English. A total of 18% of citations to RBP are self-citations, that is, authors writing in RBP citing works also published in RBP. Also, because of the increase in submission rate the submission to approval time also decreased from 130 days to about 99 days.

**CITING AUTHORS AND CITING COUNTRIES**

Another measure used by studies of journal internationalization is the citing information. BMMSS recorded only 14 citations in Google scholar received by 4 articles (Figure 2). The number of citations received by BMMSS is very small. As BMSS is only recently indexed by the ISI, its journal impact factor count and citation information is not available from the ISI database. BMMSS is also not accessible electronically and citing information could not be obtained for all articles it publishes. Hence, even though Google Scholar is a good source of information on citations, it cannot do this if the articles in the journals are not accessible electronically or the citing article is also not available on the web. When the "Bulletin" is searched in Google scholar some feedback is available but not enough to give conclusive answers about the citation status of articles in BMMSS. .

The four articles published in BMSS received a total of 14 citations and the country distribution of the citing articles is given in Table 3. Most of the citers published their works in free access journals which includes *International Journal of Mathematics and Mathematical Sciences* (7), *Journal of Inequalities and Applications* (1) and *Chaos, Solitons and Fractals* (2) and the rest were conference papers.

The results indicate that (a) there are 13 authors involved in the citing articles and they come from several countries outside of Malaysia; (b) Self-citation is evident, especially between research group members and (c) there is higher percentage of citations received from foreign authors. MJCS received a total of 88 citations for 45 out of the 164 articles that it has published between 1996 to 2007, which are all freely accessible on the Web (Table 4). MJCS is indexed by Google scholar and citation information for articles published in all issues can be obtained. That is, Google scholar will list the articles in accordance to the number of citations obtained as shown in Figure 3. The search terms posted to obtain the citation counts is site http://mjcs.fsktm.um.edu.my.

**Scholar   All articles** - **Recent articles**          Results **1 - 10** of about **18** for "**Bulletin** of the **Malaysian Mathematical Sciences Society**". (**0.05** seconds)

**All Results**

T Noiri

S Jafari

E Ekici

L References

D Bravo

[CITATION] Coalescence of difans and diwheels
D Bravo, J Rada - **Bulletin of the Malaysian Mathematical Sciences Society**
Cited by 1 - Related Articles - Web Search

[CITATION] On some difference sequence sets and their topological properties
R Çolak, M Et - **Bulletin of the Malaysian Mathematical Sciences Society**, 2005
Cited by 1 - Web Search

[CITATION] On contra-precontinuous functions
S Jafari, T Noiri - **Bulletin of the Malaysian Mathematical Sciences Society**. …, 2002
Cited by 6 - Related Articles - Web Search

[CITATION] Some properties of almost contra-precontinuous functions
T Noiri, V Popa - **Bulletin of the Malaysian Mathematical Sciences Society**. …, 2005
Cited by 3 - Related Articles - Web Search

[CITATION] Almost contra-precontinuous functions
E Ekici - **Bulletin of the Malaysian Mathematical Sciences Society**. …, 2004
Cited by 3 - Related Articles - Web Search

Figure 2: Citation information of articles in BMMSS in Google Scholar

The first article listed in Figure 3 was cited by 7 articles. Clicking "cited by 7" gives a list of the seven works that cited the article which was published in MJCS *in* 1995. The country distribution of the 88 citing works is shown in Table 5.

Table 3: Foreign and Malaysian citations of contributions in BMMSS

| Citing Countries | International | Malaysian | Total |
|---|---|---|---|
| Venenzuela/Venenzuela | 1 | - | 1 |
| India/Turkey | 1 | - | 1 |
| Japan | 1 | - | 1 |
| Egypt | 1 | - | 1 |
| Brazil/Denmark/Japan/Italy | 1 | - | 1 |
| Malaysia (UKM/UKM) | - | 1 | 1 |
| Malaysia (UPM/UPM) | - | 1 | 1 |
| Brazil/Denmark/Japan/Italy/South Arica | 1 | - | 1 |
| South Africa/USA/Brazil/Denmark | 2 | - | 2 |
| Malaysia (UKM/UKM) | - | 2 | 2 |
| USA/Turkey | 2 | - | 2 |
| Total | 10 (71%) | 4 (29%) | 14 (100%) |

Table 4: Articles in MJCS receiving citations in Google Scholar

| No. of papers | No. of citations received each | Total papers | Cumulative no. of citations |
|---|---|---|---|
| 1 | 7 | 7 | 7 |
| 2 | 5 | 10 | 17 |
| 4 | 4 | 16 | 33 |
| 5 | 3 | 15 | 48 |
| 7 | 2 | 14 | 62 |
| 26 | 1 | 26 | 88 |

The results reveal that 33 of the citers came from journal articles and the rest were from papers presented at conferences or chapter in a book or technical reports. There is high number of journal self-citations. Thirteen articles published in MJCS were citing articles published also in this journal and the citers are mostly Malaysians. This may indicate that MJCS has become a useful reference source for Malaysian ICT researchers. The results also show that researchers in the field of ICT prefer to present their findings at conferences. MJCS received about 47.7% citations from foreign authors affiliated to institutions from various countries. More Malaysian authors are citing articles in MJCS. Self-citation is also evident with authors citing their own works. The results also reveal that Google scholar is a useful tool to obtain citation information as it is currently free and especially efficient in handling open access scholarly publications.

| Scholar  All articles - **Recent articles** | Results **1** - **10** of about **164** from **mjcs.fsktm.um.edu.my** for . (**0.06** seconds) |
|---|---|
| **All Results** | **[PDF] Intelligent traffic lights control by fuzzy logic**<br>KK Tan, M Khalid, R Yusof - Malaysian Journal of Computer Science, 1995 - mjcs.fsktm.um.edu.my Malaysian Journal of Computer Science, Vol. 9 No. 2, December 1996, pp. 29-35<br>**...** INTELLIGENT TRAFFIC LIGHTS CONTROL BY FUZZY LOGIC ... Kok Khiang Tan, 1 Marzuki Khalid and Rubiyah Yusof Artificial Intelligence Center Universiti ...<br>**Cited by 7** - **Related Articles** - **View as HTML** - **Web Search** |
| **E Tee**<br>**S Lee**<br>**K Tan**<br>**N Selvanathan**<br>**K Zamli** | **[PDF] A Task-Oriented Software Maintenance Model**<br>MK Khan, MA Rashid, WNB Lo - Malaysian Journal of Computer Science, 1996 - mjcs.fsktm.um.edu.my Page 1. Malaysian Journal of Computer Science, Vol. 9 No. 2, December 1996, pp. 36-42 36 A TASK-ORIENTED SOFTWARE MAINTENANCE MODEL Md. Khaled Khan Centre for Computing and Mathematics Southern ...<br>**Cited by 5** - **Related Articles** - **View as HTML** - **Web Search** |
| | **[PDF] A Flexible and Reliable Distributed Multimedia System for Multimedia Information Superhighways**<br>JW Hong, TH Yun, JY Kong, YM Shin - Malaysian Journal of Computer Science, 1997 - mjcs.fsktm.um.edu.my  Multimedia applications are being developed and used for many aspects of our lives today. New high-speed, broadband networks have emerged and made the operation of these high-bandwidth requiring applications readily feasible. ...<br>**Cited by 4** - **Related Articles** - **View as HTML** - **Web Search** |
| | **[PDF] THE FUNDAMENTALS OF CASE-BASED REASONING: APPLICATION TO A BUILDING DEFECT PROBLEM**<br>S binti Abdullah - Malaysian Journal of Computer Science, 1997 - mjcs.fsktm.um.edu.my Traditional expert systems model human problem solving as a deductive process. They construct a solution by applying general rules to the description of a problem. Recently, however, it has become apparent that human experts rely ...<br>**Cited by 4** - **Related Articles** - **View as HTML** - **Web Search** |

Figure 3: Example of citations received by the first four articles in MJCS

Table 5: Country of works citing contributions in MJCS from Google Scholar

| Citing Countries | International | Malaysian | Total |
|---|---|---|---|
| Australia | 2 | - | 2 |
| Australia/Holland | 1 | - | 1 |
| Australia/Malaysia (Mimos) | 1 | - | 1 |
| Australia/Netherlands/Norway/Bangladesh | 1 | - | 1 |
| Bangladesh | 1 | - | 1 |
| Brazil | 2 | - | 2 |
| Brunei | - | 1 | 1 |
| China | 6 | - | 6 |
| France | 7 | - | 7 |
| Germany | 1 | - | 1 |
| Holland | 4 | - | 4 |
| Korea | 4 | - | 4 |
| Kuwait/Pakistan | - | 1 | 1 |
| Malaysia (MMU/UK) | - | 1 | 1 |
| Malaysia (U.Petronas) | - | 1 | 1 |
| Malaysia (U.Petronas/UPM) | - | 1 | 1 |
| Malaysia (UIA) | - | 1 | 1 |
| Malaysia (UKM) | - | 1 | 1 |
| Malaysia (UKM/UK) | - | 1 | 1 |
| Malaysia (UM) | - | 8 | 8 |
| Malaysia (UM/Korea) | - | 1 | 1 |
| Malaysia (UM/UK) | - | 1 | 1 |
| Malaysia (UPM) | - | 13 | 13 |
| Malaysia (UPM/MMU) | - | 4 | 4 |
| Malaysia (UPM/UM) | - | 1 | 1 |
| Malaysia (USM) | - | 2 | 2 |
| Malaysia (USM/UK) | - | 2 | 2 |
| Malaysia (USM/UM) | - | 2 | 2 |
| Malaysia (UTM) | - | 4 | 4 |
| Netherlands | 3 | - | 3 |
| New Zealand | 1 | - | 1 |
| Poland/UK/USA | 1 | - | 1 |
| Poland/USA/Canada/Germany | 1 | - | 1 |
| Sweden | 1 | - | 1 |
| Taiwan | 1 | - | 1 |
| UK | 1 | - | 1 |
| USA | 3 | - | 3 |
| **Total** | **42 (47.7%)** | **46 (52.3%)** | **88 (100%)** |

## CONCLUSION

An early observation made in 1997 (Zainab 1997) of 10 Malaysian journals indexed by discipline-based database had revealed that Malaysian journals did exhibit some characteristics of internationalization. This is especially so from the perspective of the international composition of the reviewers and the higher percentage of foreign papers being published. These characteristics are also indicated by the two journals being studied and more so in BMMSS. In summary, comparing two Malaysian journals, one indexed by the ISI and the other included in Scopus, reveal some interesting characteristics of these journals.

i)    Both journals have successfully maintained the regularity of publishing their volumes on time. This is an important factor as authors most certainly shun journals which have a long publishing lag. Regular and current publishing frequency would reassure authors that they would get to see their articles in print in the expected time. BMSS has included information such as the date of submission and the date of acceptance, which reflect the speed of the refereeing and acceptance to print process.

ii)   English is the main medium used by contributors. BMMSS do publish papers written in the Malay language but they are relatively few in number. Dinkel et al. (2004) found the internationalization of German journals in psychology by changing the publication language to English has resulted in an increase in the rates of articles published by foreign authors from about 14.6% to 52.7%. The rate of citations received also increased.

iii)  Both have adopted a more international composition of persons in their boards of editors as well as reviewers. However, the numbers of international personals is still relatively small and could be increased to 40%-50%. There should be more reviewers from the Asia-Pacific countries to ensure that Asian interests and contributions will be protected.

iv)   Both maintain a good balance between foreign and national authors in the production of an issue, to entice both international and national appeal for article submissions. This is especially indicated by BMMSS.

v)    In terms of ensuring access, MJCS have provided the journal freely accessible over the Web. This strategy has improved its accessibility to international scholars as well as researchers and Google Scholar has provided useful citation information. The results also indicate that being covered by the ISI or SCOPUS might improve international contributions but it does not necessarily improve citation counts to articles published.

Other strategies need to be adopted, such as publishing online and on open access.

vi) Both journals exhibit strong international characteristics, with reasonably high foreign submissions and acceptable citation counts from foreign works. This linkage between 'author internationalization and citation' is a strong indicator of internationality (Zitt and Bassecoulard 1998). Even though there is evidence of self-citation, both journals are getting citations from foreign authors and this is expected to increase in the future.

Although journals editorial practices, the distribution of authoring and citing countries are some of the measures that could be used to assess the internationalization of journals, there are other measures used, such as the impact factors related to the journals and the assessment by expert evaluators. Zitt and Bassecoulard (1998, 1999) studied the performance of journals in the field of earth, space and applied biology by comparing the distribution of authoring and citing countries with the average profiles of the discipline provided by the SCI indexes. Hence, when gauging the internalization of journals a number of measures should be applied to remove any biasness.

The author believes that the issue of internalization must be approached with caution. Internationalization will come at a price as it might result in national contributions being marginalized. A large number of national contributions are well written works but may find difficulties in gaining acceptance in international journals published in the developed countries for various reasons (Stern and Arndt 1999). Moreover, Nieminen et al. (2006) dispel the belief that articles published in international main stream journals are flawless. Nieminen's group evaluated all research articles published in four main stream psychiatric journals and found that about 34.6% failed to state research questions or hypotheses, 25% of the articles were difficult to read due to unclear definition of the primary response or outcome variables, 17% did not report sample size and found that citations to articles were not related to the quality of reporting. West and McIlwaine (2002) also found no correlation between the numbers of citations accrued and expert ratings of article quality in the field of addiction studies. Therefore, works published in prestige journals are not without flaws. The findings of national researches may be of value to Malaysians or those in the Asia Pacific region, which may be of less interest to the economically strong countries. The possible marginalization of national contributions is validated by Tompkins, Ko and Donovan (2001) who studied all articles published in 1983, 1988, 1993 and 1998 in 5 US and 1 UK surgical journals. The result was based upon reviewing 4,868 articles in the US journals and 1,380 articles in the British journals. The study found in the US journals, there was an increase in the total of British articles by 58.0% and the percentage of US articles decreased from 87.5% to 68.8%. Alternatively, in the British journals, the percentage of British articles also decreased from 74.8% to 47%. In all journals there was an increase in European and Asian authors. The decrease in the number of national articles, have resulted in a decrease in government funding in both countries in these disciplines. This characteristic is shown in BMMSS where there is a decrease in national articles and an increase in articles from other countries. In 2007 no Malaysian articles was published and this is a cause for concern as national funding is usually used to finance the journal.  In the quest for internationalization and in the environment of high printing and publishing cost, to give more leeway to foreign submissions will mean a decrease in the availability of space for national authors, unless the journal turns to a fully electronic platform whereby in such a situation the number of issues published or the number of articles per issue could be increased as the constraint of page cost is no longer a deterrent to publish more and frequently. This is perhaps the most viable solution for Malaysian journals which aspire for internationalization and yet safeguard national interests.